\documentstyle[amssymb]{article}

\input{tcilatex}
\begin{document}

\title{The Inertial Polarization Principle:\\
The Mechanism Underlying Sonoluminescence ?}
\author{Marcelo Schiffer\thanks{%
On leave of absence from Campinas State University} \\
The College of Judea and Samaria\\
Ariel, 44837, Israel}
\maketitle

\begin{abstract}
In this paper we put forward a mechanism in \ which imploding shock waves
emit electromagnetic radiation in the spectral region $\lambda _{0}\cong
2\pi R_{0}.$, where R$_{0}$ is the radius of the shock by the time it is
first formed.\ The mechanism relies on three different pieces of Physics:
Maxwell's equations, the existence of corrugation instabilities of imploding
shock waves and, last but not least, the{\em \ Inertial Polarization
Principle}. The principle is extensively discussed: how it emerges from very
elementary physics and finds experimental support in shock waves propagating
in water. The spectrum of\ the emitted light is obtained and depends upon
two free parameters, the amplitude of the instabilities and the cut-off $%
R_{\max }$, the shocks' spatial extension. The spectral intensity is
determined by the former , but its shape turns out to have only a mild
dependence on the latter, in the region of physical interest. The matching
with the observed spectrum $\ $requires a fine tuning of the perturbation
amplitude $\varepsilon \sim 10^{-14},${\em indicating a quantum mechanical
origin}{\it .} Indeed, we support this conjecture with an order of magnitude
estimative. The Inertial Polarization Principle clues the resolution of the
noble gas puzzle in SL.
\end{abstract}

\begin{flushleft}
\end{flushleft}

\newpage

\section*{The Inertial Polarization Principle}

In this paper we put forward a mechanism responsible for transducing the
kinetic energy stored in an imploding spherical shock wave into
electromagnetic radiation, which is based solely upon Maxwell's equations,
the existence of very small instabilities away from the spherically
symmetric flow and the inertial polarization paradigm. Based on these
premisses we \ obtained the spectral intensity of the outgoing radiation.
The mechanism turns out to be so efficient that the observed energy emission
rate of $P(\lambda )\thicksim 10^{-10}Watt/nm$ calls for perturbation
amplitudes no larger than $\varepsilon =10^{-14}!$ Maxwell's equations are a
pillar of theoretical physics while inertial polarization is a consequence
of \ very elementary physics : an atom that undergoes an acceleration, say $%
a $ , develops in its interior polarized electromagnetic fields. The issue
is made clear for an observer sitting in the frame of the molecule, where he
sees inertial forces acting both upon the nucleus $F_{N}=M_{N}a$ and on the
electronic cloud $F_{e}=M_{e}a$. The gradient between these forces tends to
sag the cloud away from the nucleus, and the atom develops internal
polarization fields, say $E_{0},$ to compensate this gradient $e$ $%
E_{0}\thicksim (M_{N}-M_{e})a$\ . The role of inertial polarization remained
hitherto unnoticed only because detectable polarization fields call for
tremendous accelerations, say, $E_{0}\thicksim 1V/m$ would require $%
a\thicksim (e/M_{p})E_{0}\thicksim
10^{-2}(eV/(M_{p}c^{2}))(c^{2}/cm)\thicksim 10^{10}cm/\sec ^{2}$ which are
absent in every day life experiments. Nevertheless, there are two instances
where such large accelerations manifest: i.) in the realm of very strong
gravitational fields where inertial polarization was shown to be the working
mechanism that rescues the second law of thermodynamics from bankruptcy
(otherwise super-luminal motion of black-holes inside dielectric media would
entail a violation of the generalized second law \cite{bekenstein},\cite
{schiffer}); ii) in the realm of shock waves, because shocks are powerful
accelerators of fluid molecules: a fluid molecule that crosses the shock
undergoes a macroscopic velocity change (of the order of the fluid velocity
itself ) within a microscopic distance -- the shock width (of the order of
the mean free path for the atomic collisions \cite{landau}).

The inertial polarization principle is the single non-very-well-established
piece of physics in our recipe and we proceed by making our case for it.
Consider a planar strong shock wave propagating within a perfect gas. Let $%
v_{2}$ and $v_{1}$ represent the fluid velocity in the back and in front the
shock, respectively (~likewise, the index 2 (1) refer to physical quantities
behind (in front) the shock ). As the fluid molecules cross the shock they
experience a mean acceleration $\bar{a}=(v_{2}-v_{1})(\bar{\Delta t})$,
where $\bar{\Delta t}$ is the mean time it takes the gas to cross the
shock-width $\delta $. Clearly $\bar{\Delta t}=\delta /\bar{v}$, where $\bar{%
v}\cong (v_{1}+v_{2})/2$, is the mean velocity. Putting these pieces
together 
\begin{equation}
\bar{a}=\frac{v_{2}^{2}-v_{1}^{2}}{2\delta }
\end{equation}

For a strong shock propagating in a perfect gas \cite{landau}: 
\begin{equation}
v_{2}^{2}-v_{1}^{2}=-\frac{2\gamma }{\gamma +1}p_{2}V_{1}
\end{equation}
where $V_{1}$ is the gas' specific volume. The compression rate satisfies $%
V_{1}/V_{2}=(\gamma +1)/(\gamma -1)$\cite{landau} 
\begin{equation}
\bar{a}=-\frac{\gamma }{\gamma -1}\frac{p_{2}V_{2}}{\delta }
\end{equation}
The shock width $\delta $ is known to be of order of the mean free path for
collisions of atoms in the fluid, $\delta \approx (n\sigma )^{-1}$, where $n$
stands for the number density of atoms and $\sigma $ for the collision's
cross section. Bearing in mind that $nV=A/\mu $ where $A$ is Avogadro's
number and $\mu $ is the molecular  weight of the gas, we obtain the
colossal figure for the mean acceleration atoms experience as they cross the
shock: 
\begin{equation}
\bar{a}\approx -\frac{\gamma }{\gamma -1}6\times 10^{13}\left( \frac{p_{2}}{%
\func{atm}}\right) \left( \frac{\sigma }{10^{-16}\func{cm}^{2}}\right)
\left( \frac{\func{gram}}{\mu }\right) \func{cm}/\sec ^{2}.
\end{equation}
The mean electric polarization developed across the shock $\bar{E}%
_{0}\approx (M_{p}/e)\bar{a}$ is also sizeable 
\begin{equation}
\bar{E}_{0}\approx 6\times 10^{3}\frac{V}{\func{meter}}\frac{\gamma }{\gamma
-1}\left( \frac{p_{2}}{\func{atm}}\right) \left( \frac{\sigma }{10^{-16}%
\func{cm}^{2}}\right) \left( \frac{\func{gram}}{\mu }\right) .
\end{equation}
Unfortunately, the shock is so thin that the voltage developed across its
ends is very small 
\begin{equation}
V\thicksim \frac{\bar{a}M_{p}\delta }{e}\thicksim 1.2\times 10^{-6}Volt\frac{%
\gamma }{\gamma -1}\left( \frac{p}{atm}\right) \left( \frac{cm^{3}}{%
g\,\varrho }\right)   \label{estimative}
\end{equation}
\qquad 

Shock Polarization was first observed in the early sixties \cite{eichelberg}
for shock waves propagating inside water. Since then, both quality and range
of the measurements improved considerably \cite{harris2}. Harris \cite
{harris1,harris2} credits the effect to the fact that large pressure
gradients inside the shock results in a torque field acting upon the water
molecule causing the molecule's dipole to align. We reproduce his results
via the table:

\[
\begin{tabular}{|l|l|l|l|l|l|l|l|l|}
\hline
$p(kbar)$ & $98$ & $75$ & $74.5$ & $58$ & $54$ & $45$ & $36$ & $20$ \\ \hline
$V(mV)/p(kbar)$ & $1.\,\allowbreak 97$ & $1.\,\allowbreak 33$ & $%
0\allowbreak .\,\allowbreak 97$ & $0\allowbreak .\,\allowbreak 77$ & $%
0\allowbreak .\,\allowbreak 43$ & $0\allowbreak .\,\allowbreak 89$ & $%
0\allowbreak .\,\allowbreak 680\,$ & $0\allowbreak .\,\allowbreak 8$ \\ 
\hline
\end{tabular}
\]

The underlying Physics for a shock propagating in water is the very same as
for a gas and we infer the averaged electric potential across the shock from
eq.(\ref{estimative}) bearing in mind that: i.)\symbol{126}the compression
rate for water is of order one, therefore we go one step back in this
equation by replacing $\gamma /(\gamma -1)\rightarrow \gamma /(\gamma
+1)\simeq 1/2$); ii.)\symbol{126}the equation was obtained for a gas and for
liquids it should be regarded as the linear expansion of the function $V(p)$%
. Then it follows that $V(mV)/p(kbar)\thicksim 0.6$ , in agreement with the
lower pressure region of the experimental data. Detection of shock
polarization for non-polar fluids would vindicate the Inertial Polarization
Principle.

The acceleration field inside planar shocks is space and time independent
(and so the corresponding polarized electromagnetic fields) . Nevertheless,
planar shocks are known to develop corrugation instabilities \cite{landau},
\ small deformations of the planar geometry that detach from the shock and
propagate throughout the fluid. They correspond to the spontaneous emission
of sound from the shock. These instabilities will cause a space time
dependent acceleration field inside the shock, and by the Inertial
Polarization Principle a wiggling $6\times 10^{3}V/m$ electric field vector
that is radiated away: sound and light are emitted simultaneously , provided
the Inertial Polarization relaxation time is small enough. \ This brings to
one's mind the famous and intriguing sonoluminescence effect \cite
{putterman+barber} in which under heavy bombarding of ultra-sound waves, a
little ( $5\mu m$) bubble of air cavitating within a flask of water
undergoes a spectacular collapse, attains the supersonic regime and glows
(mainly) violet light. The effect has been around for sixty years or so ( 
\cite{firsttime1},\cite{firstime2}) and proper understanding of the problem
remains elusive. The most popular mechanism is the Bremsstrahlung from free
electrons in the gas where the ionization is caused by two successive
heating processes: first the adiabatic collapse of the bubble which is then
followed by the motion of a shock wall inside the bubble (the shock's Mach
number controls the temperature rate $T_{2}/T_{1}\sim M^{2})$ \cite
{bremsstrahlung}.

The formation of a shock wall, a collapsing spherical front of radius $%
R(t)=A(-t)^{\alpha }$ ($\alpha <1),$ happens by the time supersonic regime
is attained inside the bubble \cite{putterman+barber}. The acceleration of
the shock front surface $a(t)\ \thicksim A(-t)^{\alpha -2}$, becomes very
large at focusing $(t\rightarrow 0)$ engendering very large space and time
dependent Inertial-Polarization fields. Nevertheless, the spherical
symmetric geometry of the problem prevents these fields to be radiated away:
pursuing the present avenue seems to require some supplementary mechanism to
account for the radiation flash (a sparking mechanism was proposed \cite
{garcia1,garcia2}). Fortunately, no supplementary mechanism is needed:
numerical calculations (\cite{roberts+wu}) have shown the existence of
unstable perturbations of the collapsing shock which provide the multipole
time-dependent inertial-polarization fields that are radiated away. The
purpose of this paper is to calculate the spectral distribution of the
emitted light .

The paper is organized as follows. The following section reviews the
dynamics of imploding shocks, and the existence of unstable multipole
perturbation modes is rigorously proved. As a bonus, we obtain the energy
and the power carried away by the sound waves that detach from the shock
(corrugation instabilities). A novel semi-analytical procedure for solving
the differential equations for the perturbations is developed, which
nevertheless, is displayed in the appendix in order prevent the disruption
of the main argument line with technicalities. In section II , we obtain the
polarization fields engendered by the corrugation instabilities and show
that they act as a source term in Maxwell's equations. Then we calculate the
spectrum of the outgoing radiation. The spectrum depends on the dynamics of
the corrugation instabilities, but fortunately it is possible to obtain the
main structure of the spectrum without having to delve too deeply into the
dynamics. The intensity of the outgoing radiation turned out to be
proportional to $p\varepsilon ^{2}$ where $\varepsilon $ is the corrugation
instability amplitude and \ $p=E_{p}^{2}/(2\alpha \hbar ),$ is
Inertial-Polarization power-constant ($E_{p}$ stands for the proton's rest
energy and $\alpha $ for the fine structure constant). This constant is of
the order $\backsimeq 1.47\times 10^{16}\func{Watt}$ (!): collapsing shock
waves are the most efficient power-stations in nature, with the sole
possible exception of astrophysical objects! Agreement with the experimental
data calls for amplitudes of the order $\varepsilon \sim 10^{-13}$ or $%
\delta r\sim 10^{-19}m$! These tiny perturbations must have a quantum
mechanical origin, and we support this conjecture by an order of magnitude
estimative. Finally we suggest the resolution of the noble gas puzzle in SL.

\bigskip

\section{Dynamics of Imploding Shocks}

The non-viscous implosion of a spherical shock cannot be characterized by
any dimensional parameter . Consequently the flow admits a self-similar
symmetry. Let $R(t)=A_{i}(-t)^{\alpha }$ represent the radius of the shock
front, where $A_{i}$ and $\alpha $ are two constants and $v_{%
shock%
%
}=\alpha R(t)/t$, its implosion velocity. The self similar parameter here is 
$\xi =r/R(t)$; the surface of the shock is given by $\xi =1$.
Self-similarity constrains the form of the speed of sound, radial flow
velocity and density \cite{guderley} : 
\begin{eqnarray}
c_{2}^{2} &=&\left( \frac{\alpha r}{t}\right) ^{2}Z(\xi ) \\
v_{2} &=&\left( \frac{\alpha r}{t}\right) V(\xi ) \\
\rho _{2} &=&\rho _{0}G(\xi )  \label{variables}
\end{eqnarray}
When expressed in terms of the self similar quantities $Z,V$ and $G$, the
boundary conditions for a strong shock $\vec{n}\cdot \vec{v}_{%
shock%
%
}>>c$ read, 
\begin{equation}
G(1)=\frac{\gamma -1}{\gamma +1}\quad ,V(1)=\frac{2}{\gamma +1}\quad ,Z(1)=%
\frac{2\gamma (\gamma -1)}{\left( \gamma +1\right) ^{2}}  \label{boundary0}
\end{equation}

The equations that govern the flow are the entropy and mass conservation
laws and Euler's equation . They provide a set of non-linear coupled
equations for $G(\xi ),V(\xi )$ and $Z(\xi )$, which when solved for $Z(V)$
and $\xi (V),$ yield the pair of equations \cite{landau} 
\begin{equation}
\frac{dZ}{dV}=\frac{Z}{1-V}\left[ \frac{\left( Z-(1-V)^{2}\right) \left(
2/\alpha -(3\gamma -1)V\right) }{\left( 3V-\kappa \right) Z-V(1-V)(1/\alpha
-V)}+\gamma -1\right]   \label{dz/dv}
\end{equation}
and 
\begin{equation}
\frac{d\ln \xi }{dV}=-\frac{Z-(1-V)^{2}}{(3V-\kappa )Z-V(1-V)(1/\alpha -V)}
\label{dxi/dv}
\end{equation}
where $\kappa =2(1-\alpha )/(\alpha \gamma )$. Inspection of these equations
reveals the existence of  a singular point at $Z=(1-V)^{2}$ ( $dV/d\xi
\rightarrow \infty ?)$. Clearly, all physical quantities, and  their
derivatives must be finite across the singular point, meaning that the
conditions $(3V-\kappa )Z-V(1-V)(1/\alpha -V)=0$ and $Z=(1-V)^{2}$ are
simultaneous to each other at this point, such as to keep their ratio
finite. Call $V_{c}(\alpha ),Z_{c}(\alpha )$ the solution of this pair of
algebraic equations. The parameter $\alpha $ is obtained by numerically
integrating $Z(V)$ from $V=V(1)$ to $V_{c}$ for different values of $\alpha $
until the matching $Z(V_{c}(\alpha ))$ = $Z_{c}(\alpha )$ is obtained. The
good values for $\alpha $ are $0.688376/0.71717$ for a monatomic/diatomic
gas. The limit $t\rightarrow 0_{-}$ corresponds to the shock's focusing
time, after which the shock reflects and reexpands$.$ For latter reference,
\ we mention the asymptotic behavior $V\sim \xi ^{-1/\alpha }$ as $\xi
\rightarrow \infty $ \cite{landau}.

We are seeking now perturbations away from this flow. Let $\delta =\delta
\rho /\rho $ be the contrast function and $\delta \vec{v}$ the velocity
fluctuation. The latter can be decomposed into its normal and perpendicular
components $\delta v_{n}=\vec{n}\cdot \vec{v}$ , $\delta \vec{v}_{\perp
}=\delta \vec{v}-\delta v_{n}\vec{n}$.

The linearized mass \ and entropy conservation equations read 
\begin{equation}
\left( \frac{\partial }{\partial t}+v\frac{\partial }{\partial r}\right)
\delta +\delta v_{n}\frac{\partial \ln \rho }{\partial r}+\vec{\nabla}\cdot
\delta \vec{v}=0  \label{cont'}
\end{equation}
\begin{equation}
\left( \frac{\partial }{\partial t}+v\frac{\partial }{\partial r}\right)
\delta s+\delta v_{n}\frac{\partial s}{\partial r}=0  \label{entropy'}
\end{equation}
while perturbing Euler's equation yields 
\begin{equation}
\left( \frac{\partial }{\partial t}+v\frac{\partial }{\partial r}\right)
\delta \vec{v}+\delta v_{n}\frac{\partial v}{\partial r}\vec{n}+\frac{v}{r}%
\delta \vec{v}_{\perp }=\frac{\delta \vec{\nabla}p-\vec{\nabla}\delta p}{%
\rho }  \label{euler'}
\end{equation}
Next, we introduce the self-similar ansatz 
\begin{eqnarray}
\delta v_{n} &=&\varepsilon \frac{\alpha r}{t_{0}}\left( \frac{t}{t_{0}}%
\right) ^{\alpha \beta -1}(1-V)\Phi (\xi )Y_{lm}(\theta ,\phi )
\label{ansatz} \\
\delta \vec{v}_{\perp } &=&\varepsilon \frac{\alpha r}{t_{0}}\left( \frac{t}{%
t_{0}}\right) ^{\alpha \beta -1}\tau (\xi )\left( r\vec{\nabla}\right)
Y_{lm}(\theta ,\phi )  \nonumber \\
\delta &=&\varepsilon \left( \frac{t}{t_{0}}\right) ^{\alpha \beta }\Delta
(\xi )Y_{lm}(\theta ,\phi )  \nonumber \\
\delta s &=&\varepsilon c_{p}\left( \frac{t}{t_{0}}\right) ^{\alpha \beta
}\sigma (\xi )Y_{lm}(\theta ,\phi )  \nonumber
\end{eqnarray}
where $c_{p}$ is the specific heat of the gas , $t_{0}$ is shock formation
time and $\varepsilon $ the amplitude of the perturbation at this moment.
After some tedious algebra we translate the previous equations in terms of
the self-similar quantities. The mass and entropy conservation yield (\ref
{cont'},\cite{entropy'}) 
\begin{equation}
(1-V)\xi \left( \Delta ^{\prime }-\Phi ^{\prime }\right) =\beta \Delta
+3\Phi -l(l+1)\tau  \label{contss}
\end{equation}
\begin{equation}
(1-V)\xi \sigma ^{\prime }=\beta \sigma -\kappa \Phi  \label{entropyss}
\end{equation}
where $\kappa =2(1-\alpha )/(\alpha \gamma )$. The projection of Euler's
equation (\ref{euler'}) into the perpendicular direction yields a compact
form 
\begin{equation}
(1-V)\xi \tau ^{\prime }=(2V+\beta -\frac{1}{\alpha })\tau +Z(\Delta +\sigma
)\,.
\end{equation}
but the normal projection gives a more cumbersome expression 
\begin{equation}
(1-V)\xi \left( (1-V)^{2}\Phi ^{\prime }-Z(\Delta ^{\prime }+\sigma ^{\prime
})\right) =
\end{equation}
\[
=\left[ (1-V)^{2}(2V+2\xi V^{\prime }+\beta -\frac{1}{\alpha })\right] \Phi
+Z((\gamma -1)\Delta +\gamma \sigma )\left( 3V+\xi V^{\prime }-\kappa
\right) 
\]
Equation (\ref{contss}) suggests the definition of a new dynamical variable $%
\Pi =\Delta -\Phi $. We display these equations in matrix form 
\begin{equation}
\frac{d}{dV}|Y(V)\rangle ={\cal M}(V)|Y(V)\rangle \quad ;\quad 0\leq V\leq
V(1)\equiv V_{1}  \label{difequation}
\end{equation}
where $|X(V)\rangle =(\phi (V),\tau (V),\pi (V),\sigma (V))$, ; $\left|
X(V)\right\rangle =\exp [\beta \int_{V_{1}}^{V}m(V)dV]|Y(V)\rangle $ and,
furthermore 
\begin{equation}
{\cal M}(V)=m(V)\left( 
\begin{array}{cccc}
P(V)\phi _{1}(V) & P(V)\phi _{2}(V) & P(V)\phi _{3}(V) & P(V)\phi _{4}(V) \\ 
Z & 2V-\frac{1}{\alpha } & Z & Z \\ 
3+\beta & -l(l+1) & 0 & 0 \\ 
-\kappa & 0 & 0 & 0
\end{array}
\right)  \label{difmatrix}
\end{equation}
with 
\begin{equation}
m(V)=\frac{1}{1-V}\frac{d\ln \xi }{dV}\quad ;P(V)=\frac{1}{(1-V)^{2}-Z}
\end{equation}
\begin{eqnarray}
\phi _{1}(V) &=&Z[5-2/\alpha +2\beta +(\gamma
-1)(3V+dV)]+(1-V)^{2}[-1/\alpha +2V+2dV] \\
\phi _{2}(V) &=&-Z\quad l(l+1)  \nonumber \\
\phi _{3}(V) &=&Z[(\gamma -1)(3V+dV-\kappa )+\beta ) \\
\phi _{4}(V) &=&Z[\gamma (3V+dV-\kappa )+\beta ]  \nonumber  \label{phin}
\end{eqnarray}
where $dV(V)\equiv \xi V^{\prime }$. Clearly this set of differential
equations possess a regular singular point when $Z-(1-V)^{2}=0$, that is to
say, at $V_{c\text{. }}$The limit $V\rightarrow 0\quad (\xi \rightarrow
\infty $ , $m(V)\rightarrow -\alpha /V,P(V)\rightarrow 1;\phi
_{1}\rightarrow -1/\alpha ;\phi _{2,3,4}\rightarrow 0)$, reveals an
additional singularity 
\begin{equation}
\frac{d}{dV}|Y(V)\rangle \approx \frac{1}{V}\left( 
\begin{array}{cccc}
1 & 0 & 0 & 0 \\ 
0 & 1 & 0 & 0 \\ 
-\alpha (3+\beta ) & \alpha l(l+1) & 0 & 0 \\ 
\alpha \kappa & 0 & 0 & 0
\end{array}
\right) |Y(V)\rangle  \label{asymptotic}
\end{equation}
The matrix on the right-hand-side of this equation defines an eigenvalue
problem whose solution 
\begin{eqnarray}
\lambda _{1,2} &=&0\quad \rightarrow \left\{ 
\begin{array}{l}
|\theta _{1}>=(0,0,1,0) \\ 
|\theta _{2}>=(0,0,0,1)
\end{array}
\right.  \nonumber \\
\lambda _{3,4} &=&1\rightarrow \left\{ 
\begin{array}{l}
|\theta _{3}>=(l(l+1),3+\beta ,0,\alpha \kappa l(l+1)) \\ 
|\theta _{4}>=(0,1,\alpha l(l+1),0)
\end{array}
\right.
\end{eqnarray}
yields the asymptotic form 
\begin{equation}
|X(V)\rangle \approx V^{-\alpha \beta }[(a_{1}|\theta _{1}>+a_{2}|\theta
_{2}>)+V(a_{3}|\theta _{3}>+a_{4}|\theta _{4}>)]\quad ;\quad V\rightarrow 0,
\label{v=0}
\end{equation}
where ${a_{n}}$ are integration constants. Asymptotically regular fields
require $\Re (\beta )\leq 0$ (except for the the particular mode $%
a_{1}=a_{2}=0$ which calls for a less stringent condition $\Re (\beta )\leq
1/\alpha $ ). A further constraint on $\beta $ arises from energetic
considerations. The energy of a polytropic gas is 
\begin{equation}
E=\int \rho \lbrack v^{2}+\frac{c^{2}}{\gamma (\gamma -1)}]dV.
\end{equation}
The lowest order contribution ( in the perturbation parameter $\epsilon $ )
to the energy stored in the perturbed-shock is the second order expression 
\begin{equation}
\delta E_{l}(t)=\int \delta \rho \lbrack v\delta v_{n}+\frac{\delta c^{2}}{%
\gamma (\gamma -1)}]4\pi r^{2}dr,
\end{equation}
or after some algebra 
\begin{equation}
\delta E_{l}(t)=4\pi \alpha ^{2}\varepsilon ^{2}\rho _{0}R_{0}^{5}\frac{%
t_{{}}^{2\alpha \beta +5\alpha -2}}{t_{0}^{\alpha (2\beta +5)}}C_{l},
\end{equation}
where 
\begin{equation}
C_{l}=\int_{1}^{\xi _{c}}G(\xi )[\Phi (\xi )+\Pi (\xi )][V(1-V)\Phi +\frac{Z%
}{\gamma (\gamma -1)}(\gamma \sigma (\xi )+(\gamma -1)(\Phi (\xi )+\Pi (\xi
))]\xi ^{4}d\xi
\end{equation}
and $R_{0}$ stands for the radius of the shock by the time it is first
formed $t_{0}$. Note that for $\xi >>\symbol{126}1\,\ ,\Phi (\xi )+\Pi (\xi
)\sim V^{-\alpha \beta }\sim \xi ^{\beta },G(\xi )\sim $ const: the integral
diverges as $\xi ^{5+2\beta }$, vindicating the introduction of the cut off $%
\xi _{c},$ which represents the boundary of the self-similarity solution.
Clearly, this energy has to remain finite at any time and at focusing it
requires that $1/\alpha -2.5\leq $ $\func{Re}(\beta )\leq 0$. In the
appendix we develop a semi-analytical method for solving eq.(\ref
{difequation}) and obtaining the correspondent spectrum for $\beta _{l,n}$.
In consonance with previous numerical calculations (\cite{roberts+wu}) \ we
confirm that $\beta $ lies in this interval. By the way, the most unstable
modes are shown to lie $\ $in the interval$.5+1/\alpha <\func{Re}(\beta
)<-2.5+3/(2\alpha )$ , even for very large values of $l$. For these modes,
the energy emission rate

\begin{equation}
P_{l}(t)=4\pi (2\alpha \beta +5\alpha -2)\alpha ^{2}\varepsilon ^{2}\rho
_{0}R_{0}^{5}\frac{t^{2\alpha \beta +5\alpha -3}}{t_{0}^{\alpha (2\beta +5)}}%
C_{l}
\end{equation}
diverges. This means that, in analogy with the corrugation instabilities in
planar shocks, a burst of sound is emitted at the focusing. The total energy
carried away during the shock-collapse is 
\begin{equation}
E_{sound}=\sum_{l=1}\delta E_{l}(t_{0})=\frac{4\pi \alpha ^{2}\varepsilon
^{2}\rho _{0}R_{0}^{5}}{t_{0}^{2}}C  \label{sound}
\end{equation}
\qquad \qquad where we defined $C=\func{Re}[\sum_{l=1,\beta }C_{l}(\beta )].$

\qquad \bigskip

\section{Inertial Polarization At Work}

As discussed already, electromagnetic bounded systems whose constituents
have sizeable mass differences, say $\Delta M$ , and which are subjected to
a strong acceleration field $d\vec{v}/dt$ engender polarization fields $\vec{%
E}_{0},\vec{B}_{0}$ that tend to restore the balance \ between
electromagnetic and inertial forces. Clearly, these polarization fields
satisfy 
\begin{equation}
\Delta M\frac{d\vec{v}}{dt}=Ze\left( {\vec{E}}_{0}+\frac{\vec{v}}{c}\times {%
\vec{B}}_{0}\right) 
\end{equation}
where $A$ and $Z$ correspond to the atomic and proton numbers and $e$ is the
electronic charge. Clearly, $\Delta M\approx AM_{p}$, where $M_{p}$ is the
proton mass. Defining a polarized potential-vector $(\Phi _{0},\vec{A}_{0})$
in the usual way, allows us to write the balance equation in the form 
\begin{equation}
\left[ \frac{\partial \vec{v}}{\partial t}-\vec{v}\times (\vec{\nabla}\times 
\vec{v})+\vec{\nabla}\frac{v^{2}}{2}\right] =-\frac{Ze}{AM_{p}c}\left[ \frac{%
\partial \vec{A}_{0}}{\partial t}-\vec{v}\times (\vec{\nabla}\times \vec{A}%
_{0})+\vec{\nabla}(c\Phi _{0})\right] 
\end{equation}
that suggests the identification $\vec{A}_{0}\rightarrow -\frac{AM_{p}c}{Ze}%
\vec{v}$ and $\Phi _{0}\rightarrow -\frac{AM_{p}}{Ze}v^{2}/2$ . Other
possible identifications exist, but they are  gauge equivalent. The
corresponding polarization fields are 
\begin{equation}
\vec{E}_{0}=\frac{AM_{p}}{Ze}\left[ \frac{\partial \vec{v}}{\partial t}+\vec{%
\nabla}\frac{v^{2}}{2}\right] \quad ;\vec{B}_{0}=-\frac{AM_{p}c}{Ze}\vec{%
\nabla}\times \vec{v}
\end{equation}
The time varying inertial-polarization fields engender the \ radiation
fields $\vec{E},\vec{B}$ and their superposition must satisfy the sourceless
Maxwell's equations: 
\begin{eqnarray}
\vec{\nabla}\cdot (\vec{E}+\vec{E}_{0})=0 &\rightarrow &\vec{\nabla}\cdot 
\vec{E}=4\pi \varrho _{eff}  \nonumber \\
\vec{\nabla}\cdot (\vec{B}+\vec{B}_{0})=0 &\rightarrow &\vec{\nabla}\cdot 
\vec{B}=0  \nonumber \\
\vec{\nabla}\times (\vec{E}+\vec{E}_{0})+\frac{1}{c}\frac{\partial }{%
\partial t}(\vec{B}+\vec{B}_{0})=0 &\rightarrow &\vec{\nabla}\times \vec{E}+%
\frac{1}{c}\frac{\partial \vec{B}}{\partial t}=0 \\
\vec{\nabla}\times (\vec{B}+\vec{B}_{0})-\frac{1}{c}\frac{\partial }{%
\partial t}(\vec{E}+\vec{E}_{0})=0 &\rightarrow &\vec{\nabla}\times \vec{B}-%
\frac{1}{c}\frac{\partial \vec{E}}{\partial t}=\frac{4\pi }{c}(%
\overrightarrow{J}_{eff}+\overrightarrow{j}_{eff})  \nonumber
\end{eqnarray}
with 
\begin{eqnarray*}
\varrho _{eff} &=&-\frac{AM_{p}}{4\pi Ze}\left[ \frac{\partial \vec{\nabla}%
\cdot \vec{v}}{\partial t}+\nabla ^{2}\frac{v^{2}}{2}\right]  \\
\overrightarrow{j}_{eff} &=&\frac{AM_{p}}{4\pi Ze}\left[ \frac{\partial ^{2}%
\vec{v}}{\partial t^{2}}+\frac{1}{2}\frac{\partial }{\partial t}(\vec{\nabla}%
v^{2})\right] 
\end{eqnarray*}
clearly satisfying the conservation equation $\partial \varrho
_{eff}/\partial t+\overrightarrow{\triangledown }\cdot \overrightarrow{j}%
_{eff}=0$ \ and 
\[
\overrightarrow{J}=\frac{AM_{p}c^{2}}{4\pi eZ}\vec{\nabla}\times \vec{\nabla}%
\times \vec{v}
\]

For non-relativistic flows\ $|j^{\mu }|/|J^{\mu }|\sim (L/T)^{2}/c^{2}\sim
v^{2}/c^{2}\ $, and the field equations reduce to 
\begin{eqnarray}
\vec{\nabla}\cdot \vec{E} &=&0  \label{maxwell} \\
\vec{\nabla}\cdot \vec{B} &=&0  \nonumber \\
\vec{\nabla}\times \vec{E}+\frac{1}{c}\frac{\partial \vec{B}}{\partial t}
&=&0  \nonumber \\
\vec{\nabla}\times \vec{B}-\frac{1}{c}\frac{\partial \vec{E}}{\partial t}
&=&\eta \vec{\nabla}\times \vec{\nabla}\times \vec{v}  \nonumber
\end{eqnarray}
where $\eta =AM_{p}c/Ze$. Next we expand $\left( 
\begin{array}{c}
\vec{E} \\ 
\vec{B}
\end{array}
\right) =\sum_{n=1}^{3}\left( 
\begin{array}{c}
E_{n} \\ 
B_{n}
\end{array}
\right) \vec{e}_{n}$ where $\vec{e}_{n}$ is the familiar vector basis \cite
{jackson}: 
\begin{equation}
{\cal E}=\left( \vec{e}_{1},\vec{e}_{2},\vec{e}_{3}\right) =\left( \vec{n}%
Y_{lm}(\theta ,\phi ),(r\vec{\nabla})Y_{lm}(\theta ,\phi ),(\vec{r}\times 
\vec{\nabla})Y_{lm}(\theta ,\phi )\right) 
\end{equation}

For latter reference we mention the following identities: 
\begin{equation}
\vec{\nabla}\cdot {\cal E}=\frac{Y_{lm}}{r}(2,-l(l+1),0);\vec{\nabla}\times 
{\cal E}=\frac{1}{r}(-\vec{e}_{3},\vec{e}_{3},-\vec{e}_{2}-l(l+1)\vec{e}_{1})
\end{equation}

The unperturbed flow is rotation free and the leading contribution to
Maxwell's equations [eq. (\ref{maxwell})] comes from the perturbed  flow $%
\delta \vec{v}=\delta v_{n}\vec{e}_{1}+\delta v_{\perp }\vec{e}_{2}$,

\begin{equation}
\frac{\partial (r^{2}B_{1})}{\partial r}-l(l+1)B_{2}r=0  \label{field1}
\end{equation}

\begin{equation}
\frac{\partial (r^{2}E_{1})}{\partial r}-l(l+1)E_{2}r=0  \label{field2}
\end{equation}

\begin{equation}
\frac{r}{c}\dot{B}_{1}-l(l+1)E_{3}=0  \label{field3}
\end{equation}

\begin{equation}
\frac{r}{c}\dot{B}_{2}-\frac{\partial (rE_{3})}{\partial r}=0  \label{field4}
\end{equation}

\begin{equation}
\frac{r}{c}\dot{B}_{3}-E_{1}+\frac{\partial (rE_{2})}{\partial r}=0
\label{field5}
\end{equation}

\begin{equation}
\frac{r}{c}\dot{E}_{1}+l(l+1)B_{3}=\eta l(l+1)f(r,t)  \label{field6}
\end{equation}

\begin{equation}
\frac{r}{c}\dot{E}_{2}+\frac{\partial (rB_{3})}{\partial r}=\eta \frac{%
\partial (rf)}{\partial r}  \label{field7}
\end{equation}

\begin{equation}
-\frac{r}{c}\dot{E}_{3}-B_{1}+\frac{\partial (rB_{2})}{\partial r}=0
\label{field8}
\end{equation}
where $f(r,t)=\frac{\partial \delta v_{\perp }}{\partial r}+\frac{\delta
v_{\perp }-\delta v_{n}}{r}$. Notice that $B_{2},B_{1\text{ }}$and$\ E_{3}$
are independent of the source term, and are \ taken to vanish identically.
The other mode is 
\begin{equation}
\vec{E}=E_{1}\vec{e}_{1}+E_{2}\vec{e}_{2}\quad ;\vec{B}=B_{3}\vec{e}_{3}.
\label{polarized}
\end{equation}
Averaging the Poynting vector 
\begin{equation}
\vec{S}=\frac{c}{8\pi }(\vec{E}\times \vec{B^{\ast }})=\frac{cB_{3}^{\ast }}{%
8\pi }\left[ -rE_{1}(Y\vec{\nabla}Y^{\ast })+r^{2}E_{2}(\vec{\nabla}Y\cdot 
\vec{\nabla}Y^{\ast })\vec{n}\right] ,
\end{equation}
over all directions gives the radial energy flux 
\begin{equation}
S_{r}=\frac{cl(l+1)}{8\pi }\Re (E_{2}B_{3}^{\ast }).
\end{equation}
The corresponding spectral intensity is

\begin{equation}
I_{l}(\omega )=\frac{cr^{2}l(l+1)}{2}\left| E_{2}(\omega )B_{3}^{\ast
}(\omega )\right|
\end{equation}

\bigskip We obtain the wave equation for $\Lambda \equiv E_{1}(\omega )r$ by
combining eqs.(\ref{field1})-(\ref{field8})

\begin{equation}
(\nabla _{r}^{2}+k^{2})\Lambda (\omega )=ik\eta l(l+1)f(\omega ,r)
\end{equation}
and in terms of $\Lambda ,$ the spectral intensity reads

\begin{equation}
I_{l}(\omega )=\frac{\omega }{2l(l+1)}\left| r\Lambda (\omega )\frac{%
\partial (r\Lambda ^{\ast }(\omega ))}{\partial r}\right|
\end{equation}

The wave equation is solved through the Green's function method in the
region away from the near zone:

\begin{equation}
\Lambda (\omega )=ikh_{l}^{(1)}(kr)\int [-ik\eta l(l+1)f(\omega ,r\prime
)]j_{l}(kr\prime )r^{\prime 2}dr\prime .
\end{equation}

In the \ radiation zone, $\Lambda (\omega )$ reduces to :

\begin{equation}
\Lambda (\omega )r\approx -e^{ikr}(-i)^{l+1}k\eta l(l+1)\int f(\omega
,r\prime )j_{l}(kr\prime )r\prime ^{2}dr\prime
\end{equation}

Putting these pieces together,

\begin{equation}
I_{l}(\omega )=\frac{1}{2}c\eta ^{2}k^{4}l(l+1)\left| A_{l}(k)\right| ^{2}
\end{equation}
with 
\begin{equation}
A_{l}(k)=\int \int f(r,t)e^{-i\omega t}j_{l}(kr)r^{2}drdt
\end{equation}

The function $f(r,t)$ can be expressed in terms of the fluctuation functions
[eqs.(\ref{ansatz})], 
\begin{equation}
f(r,t)=\frac{\alpha \varepsilon }{t_{0}}\left( \frac{t}{t_{0}}\right)
^{\alpha \beta -1}[\xi \tau ^{\prime }(\xi )+2\tau (\xi )-(1-V(\xi ))\Phi
(\xi ).
\end{equation}
Calling $x=kr$ and performing a change of integration variables we obtain
radiation emission rate per wave-length $\lambda $: 
\begin{equation}
P_{l}(\lambda )=\frac{p}{\lambda }\varepsilon ^{2}\alpha ^{2}l(l+1)\left|
W_{l}(k)\right| ^{2}  \label{power}
\end{equation}
with 
\begin{equation}
W_{l}(k)=\int_{0}^{\infty }j_{l}(x)\quad x^{2}dx\int_{0}^{1}[\xi \tau
^{\prime }+2\tau -(1-V)\Phi ]y^{\alpha \beta -1}\exp [-iQy]dy,
\end{equation}
where $Q\equiv kR_{0}(t_{0}/R_{0}c)$ and $p\equiv c^{3}\eta ^{2}/2$ .
According to Barber ($\cite{barber94}$) the ratio $\alpha R_{0}/t_{0}=c_{0}$%
, the speed of sound, and $Q=\alpha \ kR_{0}(c_{0}/c)\sim 10^{-5}(kR_{0}).$
The asymptotic behavior given by eq.(\ref{v=0}) and the fact that $\
V\varpropto \xi ^{-1/\alpha }$ suggests the expansion:

\begin{equation}
\lbrack \xi \tau ^{\prime }+2\tau -(1-V)\Phi ]=\sum_{n=1}b_{n}\xi ^{\beta
-n/\alpha }=\sum_{n=1}b_{n}\left( \frac{x}{kR_{0}}\right) ^{\beta -n/\alpha
}y^{n-\alpha \beta }
\end{equation}
where the coefficients $b_{n}$ are determined by the dynamics of
perturbations. Note that the sum does not contain the $n=0$ term because the
leading term of the series [see again eq.(\ref{v=0} )] for the velocity
components $\Phi ,\tau $ is $V^{1-\alpha \beta }.$ \ Therefore,

\begin{equation}
W_{l}(k)=\sum_{n=1}b_{n}(kR_{0})^{n/\alpha -\beta }\int_{0}^{1}y^{n-1}\exp
[-iQy]dy\int_{0}^{kR_{\max }}j_{l}(x)\quad x^{2+\beta -n/.\alpha }dx.
\label{window}
\end{equation}

The cutoff $kR_{\max \text{ }}$in the x-integral was introduced because the
shock does not extend beyond $R_{max}$, the ambient radius of the bubble.
For $Q<<1$ we might transform this expression into

\begin{equation}
W_{l}(k)=(kR_{0})^{-\beta }\sum_{n=1}\frac{b_{n}}{n}\left[
(kR_{0})^{n/\alpha }\int_{kR_{0}}^{kR_{\max }}j_{l}(x)\quad x^{2+\beta
-n/.\alpha }dx+\int_{0}^{kR_{\max }}j_{l}(x)\quad x^{2+\beta }dx\right]
\label{2integrals}
\end{equation}

The detailed form of the spectrum requires a full knowledge of $b_{n}$, that
is to say, dynamics of the fluctuations must be specified (this can be done
analytically by using the method developed in the appendix). Fortunately,
the major features of the spectrum can be obtained without delving into the
differential equations. For instance, in the region where $kR_{\max }<1$ we
can approximate $j_{l}(x)\simeq (2x)^{l}l!/(2l+1)!$ and then

\begin{equation}
W_{l}(k)\simeq \frac{2^{l}l!}{(2l+1)!}\left( kR_{0}\right) ^{l+3}\sum_{n=1}%
\frac{b_{n}}{n}\left\{ \frac{1}{l+3+\beta -n/\alpha }\left[ \left( \frac{%
R_{\max }}{R_{0}}\right) ^{l+3+\beta -n/\alpha }-1\right] -\frac{1}{%
(l+3+\beta )}\left( \frac{R_{\max }}{R_{0}}\right) ^{l+\beta +3}\right\}
\end{equation}

In the other end of the spectrum $kR_{0}>1$, taking the asymptotic
expression $\ j_{l}(x)\ \approx 1/x\sin (x-l\pi /2)$ is justified, either
because in the first integral the integration variable $\ x>1$ or because\
in the second integral the measure $x^{2+\beta }$ (with $2+\beta >1$)
ensures that important contributions to the integral comes from the large
arguments. Thus,

\begin{equation}
W_{l}(k)\simeq \left( kR_{0}\right) \left( \frac{R_{\max }}{R_{0}}\right)
^{\beta +1}\sum_{n=1}\frac{b_{n}}{n}\left[ f(\beta ;kR_{\max })+\left( \frac{%
R_{0}}{R_{\max }}\right) ^{n/\alpha }f(\beta -n/\alpha ;kR_{\max })-f(\beta
-n/\alpha ;kR_{0})\right] ;
\end{equation}
where

\begin{equation}
f(\beta ;x)=\func{Im}\left[ e^{-il\pi /2}\sum_{m=0}^{\infty }\frac{(ix)^{m+1}%
}{(m+\beta +2-n/\alpha )m!}\right] .  \label{f(x)}
\end{equation}

The dominant power low contribution to $W_{l}(k)\ $ in the region $kR_{0}>1$
comes from the linear term $\left( kR_{0}\right) $ \ because the series $%
f(\beta ;x)$ behaves nearly like $\sin (x)$, for $x>1.$ Taking the following
figures $R_{\max }\sim 5\mu m$ , the ambient radius of the bubble and $%
R_{0}\sim 0.15\mu m,$ (we shall explain in a moment) and defining $\lambda
_{0}=2\pi R_{0},$ we display our asymptotic expressions in the form

\begin{equation}
P_{l}(\lambda )\sim p\varepsilon ^{2}\left\{ 
\begin{array}{c}
A_{l}\lambda ^{-1}\left( \lambda _{0}/\lambda \right) ^{2l+6};\lambda
>>\lambda _{0} \\ 
\lambda _{0}^{2}/\lambda ^{3}g_{l}(\lambda );\lambda <\lambda _{0}
\end{array}
\right.  \label{powerlimit}
\end{equation}
where 
\begin{equation}
A_{l}=l(l+1)\left| \alpha \frac{2^{l}l!}{(2l+1)!}\sum_{n=1}\frac{b_{n}}{n}%
\left\{ \frac{1}{l+3+\beta -n/\alpha }\left[ \left( \frac{R_{\max }}{R_{0}}%
\right) ^{l+3+\beta -n/\alpha }-1\right] -\frac{1}{(l+3+\beta )}\left( \frac{%
R_{\max }}{R_{0}}\right) ^{l+\beta +3}\right\} \right| ^{2}
\end{equation}
and 
\begin{equation}
g_{l}(\lambda )=l(l+1)\left| \alpha \left( \frac{R_{\max }}{R_{0}}\right)
^{\beta +1}h_{l}(k)\right| ^{2}  \label{gl}
\end{equation}
with

\begin{equation}
h_{l}(k)=\sum_{n=1}\frac{b_{n}}{n}\left[ f(\beta ;kR_{\max })+\left( \frac{%
R_{0}}{R_{\max }}\right) ^{n/\alpha }f(\beta -n/\alpha ;kR_{\max })-f(\beta
-n/\alpha ;kR_{0})\right]
\end{equation}

The apparent divergence of $g_{l}(\lambda )$ at  large angular momenta [see
eq. (\ref{gl})] seems to endanger the present results. This worry is removed
studying the asymptotic behavior $g_{l}(\lambda )$, bearing in mind that in
this limit $\ \beta \simeq \pm il\sqrt{(\gamma -1)/(\gamma +1)},$ \cite
{wu+roberts}. This yields that $g_{l}(\lambda )\rightarrow 0$ as $%
l\rightarrow \infty $, regardless of the specific form of the dynamical
coefficients $b_{n}$ may take.

\bigskip

\section{ Assessment of the Results}

The present SL mechanism relies on very basic pieces of physics, the
existence of corrugation instabilities in spherical shocks, whose existence
is well known, Maxwell's equations and the inertial polarization paradigm.
As we had the opportunity to explain, this paradigm stems from very
elementary physics and it has\ remained hitherto unnoticed only because huge
accelerations are required for sizeable polarizations. The detection of
shock polarization in non-polar liquids would lend an undisputable status to
the inertial polarization principle . In the transduction of sound into
radiation , the flash of light must be coincident with a burst of sound
since the emission of radiation is caused by corrugation instabilities .
According to\ eq.(\ref{polarized}), only one field-mode is related to the
sonoluminescent light. This mode has a longitudinal electric field component 
$E_{1}$, and some experiment must be devised to detect it .The transversal
component $E_{2}\ $points into the direction of the vector

\[
\overrightarrow{e}_{2}=\sqrt{\frac{2l+1}{4\pi }\frac{(l-m)!}{(l+m)!}}%
e^{im\varphi }\sin \left( \theta \right) \left( imP_{l}^{m}(\cos (\theta ))%
\overrightarrow{e}_{\varphi }-P_{l}^{\prime m}(\cos (\theta ))%
\overrightarrow{e}_{\theta }\right) 
\]
and this (weird) polarization should be observed in sonoluminescent light. $%
\ $

\bigskip

Physics is seldom controlled by cut-off parameters, and we expect the
cut-off parameter $R_{max}$ ( the bubble's ambient radius) to play a
marginal role in delimiting the frequency band where light is emitted. The
main features of the spectrum should be controlled by the remaining
parameters: $R_{0},$ the radius of the shock-wave when it is first formed
and the perturbation amplitude $\varepsilon $ . Thus, $R_{0}$ should
characterize the typical wave-length of the emitted light $\lambda \approx
\lambda _{0}=2\pi R_{0}$. Our asymptotic results [eq.(\ref{powerlimit})]
confirms this feeling. Numerical and theoretical studies of the dynamics of
imploding shocks \ support the picture that the bubble collapses at the
speed of sound by the time it passes through its ambient radius as the right
criterion both for shock formation and the existence of SL (\cite{barber94}- 
\cite{wu+roberts94}). According to these investigations, at $100ps$ before
the bubble reaches its minimum size, a shock wave of initial radius $%
R_{0}=0.15\mu m$ develops: by this time the interface is imploding with 4 to
5 times the ambient speed of sound. \ With these figures, we predict the
emitted light to lie in $\lambda \approx \lambda _{0}=900nm$ spectral
region, regardless the kind of gas present in the bubble; in SL experiments
light is observed in the $\ 200nm\lesssim $ $\lambda \lesssim 800nm$
interval. According to this result, it is legitimate to infer the spectrum
in this wave-length interval through the asymptotic formula for $\lambda
\precsim \lambda _{0}$[see eq.(\ref{powerlimit})]. How does the particular
kind of gas present in the bubble impact on the the emission power? The
dependence of the emitted light upon the particular type of gas present in
the bubble stems from two different factors:

i.) different values of the adiabatic index $\gamma $ leads to a different
shock-wave and corrugation instability dynamics; ii.) different gases have
different dielectric permeability $\epsilon $.

The dielectric nature of the gas is implemented through the replacement $%
E\rightarrow D$ in the Poynting vector, which corresponds to the replacement
of $\left| W_{l}(k)\right| ^{2}$ by $\widetilde{\epsilon }(k)\left|
W_{l}(k)\right| ^{2}$ , or $g_{l}(\lambda )$\bigskip $\rightarrow $ $%
\widetilde{\epsilon }(k)g_{l}(\lambda )$\bigskip\ . Different adiabatic
indexes would cause $\ h_{l}(k)$ to change because both the spectrum of $%
\beta $ and the dynamical coefficients $b_{n}$ , depend upon $\gamma $.
These two conditions will cause a change on the shape of the function $%
g_{l}(\lambda )$\bigskip\ . Assuming that after taking these corrections
into account, the function $g_{l}(\lambda )$ still remains marginally
dependent upon the wave-length (non power law), the overall change produced
by different gases in the shape on the logartithmic representation of the
spectrum $\ln P\sim -3\ln \lambda +\ln g_{l}(\lambda )+const$ for $\lambda
\lesssim \lambda _{0}$ , is a displacement of the nearly parallel lines of
inclination $m\ \cong -3$. This behaviour is changed as we approach the $%
\lambda <<\lambda _{0}$ region because then the dielectric constant being
governed by the plasma frequency of the gas, causes the function $\ln
g_{l}(\lambda )$ to strongly depend upon $\lambda $ .

\bigskip

Infering the uncorrected spectra for transmission by the surrounding medium
observed by Hiller in SL experiments for bubbles trapping pure noble gases
bubbles at $0^{0}C$ (\cite{hiller}) we infered $m\ \cong -2.7$. For pure He, 
$m\ \cong -2.5.$ \ Inspection of the spectra shows the nearly linear
dependence for $Ar,He,He^{3},$and $Ne$. The agreement is less accurate for $%
Xe$ and $Kr,$ for reasons which are presently unclear: it might well be the
that heavier noble gases cannot be handled with the naive classical\
Inertial Polarization picture, they have too much internal structure and
must be handled with a full quantum mechanical approach. The spectrum for a
mixture of 1\% of $He$ and $N_{2}$ closely resemble the behavior of pure He$%
_{2}$(\cite{hiller}). Differences might be credited to the superimposition
of the Bremsstrahlung spectrum of free electrons of the ionized $N_{2}$ gas
in the mixture to the original spectrum, or even the effect of the
Inertial-Polarization fields upon\ these electrons.

\bigskip

Regarding now the intensity of the outgoing radiation, it is governed by the
product$\ p\varepsilon ^{2}\ $. A small $p$ would require large corrugation
instabilities, invalidating the linear regime approximations. Surprisingly, $%
p=E_{p}^{2}/(2\hbar \alpha )\backsimeq 1.47\times 10^{16}\func{Watt}$ ,
imploding shocks are fantastic power stations ! Actually, we have to worry
to have sufficiently small perturbations to fit the experimental data!
Typical power emissions are of of the order of $10^{-11}Watt/nm$ in the $%
\lambda _{0}$ region \cite{hiller}, \ calling for an amplitude $\varepsilon
\sim 10^{-12}$ or $\delta r=\varepsilon R_{0}\sim 10^{-19}m$, which being
much smaller then the nuclear dimensions can have only a quantum mechanical
origin. Now, the radius of the shock at the moment it is formed $R_{0}$ is
governed by the radius of the bubble wall $R_{b}$, \bigskip by the time it
is collapsing at 4-5 times the ambient speed of sound. The dependence of the
former on the latter is linear. In a semi-classical approach, it is to be
expected that the fluctuations on the shape of the imploding shock are also
governed by bubble wall fluctuations, $\epsilon =\delta R_{0}/R_{0}=\delta
R_{b}/R_{b}.$ The fluctuations of the bubble interface should be of the
order of the bubble's Compton wave-length $\lambda _{b}$ and $\delta
r=(R_{0}/R_{b})\lambda _{b}\sim \lambda _{p}/N$, where $\lambda _{p}\sim
10^{-15}m$ is the Compton wave-length of the proton and $N$\ is the number
of gas atoms trapped inside the bubble , $N\sim 10^{7}.$ Thus, in this
scenario $\delta r\sim 10^{-22}m$, \ which is close to the amplitude needed
to fit the observed intensity of the radiation.

\bigskip

One of the most intriguing issues in SL is beyond any doubt the noble gas
puzzle: only bubbles containing noble gas,even at very small concentrations,
glow. What can we say in this respect? \ Does our paradigm shed some light
in this direction? Here is a clue. As the bubble collapses and the attains
supersonic regime the adiabatic heating \ raises the gas temperature to $%
\sim 0.4eV$ (\cite{lofstedt}) . The gas is further heated when it crosses
the shock front, the temperature is increased by a factor $M^{4}$ . This is
more than enough to bring diatomic gases to their excited states, but not
for noble gases. The dipole contribution $\left\langle \Psi _{lmn}\left| 
\widehat{p}\right| \Psi _{lmn}\right\rangle \cdot r^{-3}$ of the excited
states being much larger than the Inertial Polarization Fields will wash
away information regarding the latter. A full quantum mechanical calculation
should resolve this issue.

There are immense challenges ahead. From the theoretical point of view, one
needs to calculate the detailed spectrum taking full account of the shock
dynamics, study the back reaction of the polarized fields upon the dynamics,
clarify whether the polarization caused by quantum mechanical transitions of
a diatomic molecule are the culprits for washing out the Inertial
Polarization fields, etc. The immediate experimental challenge is to detect
Shock Polarization in non-polar fluids. If the effect is confirmed in
non-polar fluids then it will be very hard to defuse the present
transduction mechanism\bigskip .

\subsection*{Appendix -- A semi-analytical solution of the differential
equations for the perturbed flow}

In order to solve the set of differential equations we split the matrix into
its regular and divergent parts

\begin{equation}
{\cal M}(V)=\frac{{\cal A}}{(V-V_{c})}+{\cal B}(V)
\end{equation}
with

\begin{equation}
{\cal A}=\frac{m(V_{c})}{(dP/dV)_{V_{c}}}\left( 
\begin{array}{cccc}
\phi _{1} & \phi _{2} & \phi _{3} & \phi _{4} \\ 
0 & 0 & 0 & 0 \\ 
0 & 0 & 0 & 0 \\ 
0 & 0 & 0 & 0
\end{array}
\right) ;{\cal B}=m(V)\left( 
\begin{array}{cccc}
\tilde{\phi}_{1}(V) & \tilde{\phi}_{2}(V) & \tilde{\phi}_{3}(V) & \tilde{\phi%
}_{4}(V) \\ 
Z & 2V-\frac{1}{\alpha } & Z & Z \\ 
3+\beta & -l(l+1) & 0 & 0 \\ 
-\kappa & 0 & 0 & 0
\end{array}
\right)
\end{equation}
where $\phi _{\alpha }$is a short notation for $\phi _{\alpha }(V_{c})$ and $%
\phi _{\alpha }(V)=$ $\phi _{\alpha
}(V)/P(V)-m(V_{c})/(m(V)(dP/dV)_{V_{c}}(V-Vc))$. Assuming $B(V)$ and $\left|
Y(V)\right\rangle $ regular functions at the critical point $V_{c}$ permits
the expansions $B(V)=\sum_{n}B_{n}(V-V_{c})^{n};\left| Y(V)\right\rangle
=\sum_{k}{\bf Y}_{k}(V-V_{c})^{k}$. Substitution into the differential
equation yields the recurrence formulae:

\begin{eqnarray}
{\cal A}{\bf Y}_{0} &=&0  \label{recurrence} \\
{\bf Y}_{n+1} &=&[(n+1){\cal I}-{\cal A}]^{-1}\sum_{m\leq n}{\cal B}_{n-m}%
{\bf Y}_{m}
\end{eqnarray}
The matrix ${\cal A}$ possess three distinct null-eigenvectors:

\begin{eqnarray}
Y_{0}^{(2)} &=&(-\phi _{2},\phi _{1},0,0)  \nonumber \\
Y_{0}^{(3)} &=&(-\phi _{3},0,\phi _{1},0) \\
Y_{0}^{(4)} &=&(-\phi _{4},0,0,\phi _{1}).  \nonumber
\end{eqnarray}

Associated to each one of these eigenvectors we can construct through the
recurrence relations \ $\left| Y^{(i)}(V)\right\rangle .$ The solution of
the differential equation is the linear combination $\left|
Y(V)\right\rangle =\sum_{i=2,4}c_{i}$ $\left| Y^{(i)}(V)\right\rangle .$ The
fulfillment of the boundary requires that

\begin{equation}
\left| X(V_{1})\right\rangle =\left| Y(V_{1})\right\rangle =\sum_{k}[c_{2}%
{\bf Y}_{k}^{(2)}+c_{3}{\bf Y}_{k}^{(3)}+c_{4}{\bf Y}%
_{k}^{(4)}](V_{1}-V_{c})^{k}.  \label{betaequation}
\end{equation}
\qquad \qquad \qquad\ 

This equation constitutes a set of four equations for the unknown $%
(c_{i},\beta )$, which can be solved in a perturbational approach in powers $%
(V_{1}-V_{c})$, once the state $\left| X(V_{1})\right\rangle $ is known. The
only missing piece of information is the set of boundary conditions for the
perturbed fields.

\bigskip

\subsubsection*{The boundary conditions for the perturbed flow}

Supersonic motion produces a discontinuity in the fluid flow known as a
shock wave or simply shock. Let us call $\vec{v}_{2}$ and $\rho _{2}$ the
fluid velocity and density, and $c_{2\text{ }}$the speed of sound behind the
shock, as measured in the laboratory frame(likewise, the subscript $1$
refers to the same quantities in the front of the shock). The normal to the
shock is $\vec{n}$ and its velocity in the lab frame is $\vec{v}_{%
shock%
%
}$. The discontinuities have to fulfill the following conditions at the
shock surface \cite{landau} 
\begin{equation}
\vec{n}\times \left[ \vec{v}_{1}-\vec{v}_{%
shock%
%
}\right] =\vec{n}\times \left[ \vec{v}_{2}-\vec{v}_{%
shock%
%
}\right]  \label{boundary1}
\end{equation}
\begin{equation}
\frac{\vec{n}\cdot \left[ \vec{v}_{2}-\vec{v}_{%
shock%
%
}\right] }{\vec{n}\cdot \left[ \vec{v}_{1}-\vec{v}_{%
shock%
%
}\right] }=\frac{\rho _{1}}{\rho _{2}}=\frac{\gamma -1}{\gamma +1}
\label{boundary2}
\end{equation}
\begin{equation}
c_{2}^{2}=\frac{\gamma -1}{\gamma +1}\left[ c_{1}^{2}+\frac{2\gamma }{\gamma
+1}\left( \vec{n}\cdot \vec{v}_{%
shock%
%
}\right) ^{2}\right]  \label{boundary3}
\end{equation}

In the perturbed \ flow the shock front is displaced from $\Sigma
_{0}:r-R(t)=0 $ to ($\Sigma _{0}+\delta \Sigma ):r-R(t)-\delta r(t,\theta
,\phi )=0$. The corresponding perturbed normal is $\delta \vec{n}=-\vec{%
\nabla}\delta r$, the perturbed shock velocity is $\dot{\delta r}$ while the
location of the shock itself in self-similar coordinate is $1+\delta \xi $, $%
\delta \xi =\delta r/R(t)$. Accordingly, 
\begin{eqnarray}
\delta \xi &=&\varepsilon \left( \frac{t}{t_{0}}\right) ^{\alpha \beta
}Y_{lm}(\theta ,\phi )  \nonumber \\
\delta r &=&\varepsilon R(t)\left( \frac{t}{t_{0}}\right) ^{\alpha \beta
}Y_{lm}(\theta ,\phi )  \nonumber \\
\delta \vec{n} &=&-\varepsilon \left( \frac{t}{t_{0}}\right) ^{\alpha \beta
}(R(t)\vec{\nabla})Y_{lm}(\theta ,\phi )  \nonumber \\
\delta \vec{v}_{s} &=&\varepsilon \alpha (1+\beta )R(t)\left( \frac{t}{t_{0}}%
\right) ^{\alpha \beta -1}Y_{lm}(\theta ,\phi )\vec{n}  \label{hafraot}
\end{eqnarray}
The first order corrections to the boundary conditions [eqs. (\ref{boundary1}%
)-(\ref{boundary3}) ] are: 
\begin{eqnarray}
\vec{n}\times \left[ \left( \frac{\partial v}{\partial \xi }\delta \xi
+\delta v_{n}-\delta v_{s}\right) \vec{n}+\delta \vec{v}_{\perp }\right]
+\delta \vec{n}\times \vec{n}\left( v-v_{s}\right) &=&-\vec{n}\times \delta 
\vec{v}_{s}-\delta \vec{n}\times \vec{n}\,v_{s}  \nonumber \\
\delta \rho _{2}(1)+\frac{\partial \rho _{2}}{\partial \xi }\delta \xi &=&0
\\
\vec{n}\cdot \left[ \left( \frac{\partial v}{\partial \xi }\delta \xi
+\delta v_{n}-\delta v_{s}\right) \vec{n}+\delta \vec{v}_{\perp }\right]
+\delta \vec{n}\cdot \vec{n}(v-v_{s}) &=&-\frac{\gamma -1}{\gamma +1}(\delta
v_{s}+\vec{n}\cdot \delta \vec{n}\,v_{s})  \nonumber \\
\delta c_{2}^{2}+\frac{\partial c_{2}^{2}}{\partial \xi }\delta \xi
&=&2Z(1)\,v_{s}\left( \delta \vec{n}\cdot \vec{v_{s}}+\vec{n}\cdot \delta 
\vec{v}_{s}\right)  \nonumber
\end{eqnarray}
Inserting eqs.(\ref{hafraot})-(\ref{ansatz}) into these boundary conditions,
yields 
\begin{eqnarray}
\Phi _{1} &=&\frac{\beta V_{1}-V_{1}^{^{\prime }}}{1-V_{1}}\quad ;  \nonumber
\\
\tau _{1} &=&-V_{1}\quad ;  \nonumber \\
\Delta _{1} &=&-\frac{G_{1}^{^{\prime }}}{G_{1}}=  \nonumber \\
\frac{\delta Z_{1}}{Z_{1}} &=&(2\beta -\frac{Z_{1}^{^{\prime }}}{Z_{1}})
\end{eqnarray}

or, equivalently

\begin{equation}
\left| X(V_{1})\right\rangle =\varepsilon \left( 
\begin{array}{c}
(\beta V_{1}-V_{1}^{^{\prime }})/(1-V_{1}) \\ 
-V_{1} \\ 
-(\beta +3)V_{1}/(1-V_{1}) \\ 
2\left( \beta +(1/\alpha -V_{1})/(1-V_{1}\right) )/\gamma
\end{array}
\right)  \label{boundary}
\end{equation}

\subsubsection*{Numerical Procedure}

Our procedure for resolving the spectrum of $\beta $ consists of first
fitting the unperturbed flow ($Z(V),dV(V)$) by a polynomial in $V$, from
which we extract the matrices ${\cal A}\ \ $and ${\cal B}(V)\ $ as power
series in $V$. Then through the recurrence formulae (\ref{recurence}) we
obtain the expansion coefficients ${\bf Y}_{n}$($\beta )$ up to a given
order and insert then into eq.(\ref{betaequation}), in conjunction with the
above boundary condition $\left| X(V_{1})\right\rangle $ [eq.(\ref{boundary}%
) ] . This procedure yields a polynomial equation for $\beta $ , which is
solved numerically. \ We display the results for $\gamma =7/5,l=1,2,3,4.$

\FRAME{ftbpFU}{6.4463in}{3.979in}{0pt}{\Qcb{Real and Imaginary parts of $%
\protect\beta $ for l=1,2,3,4.}}{\Qlb{beta}}{beta.eps}{}

\bigskip

$\bigskip $

$\bigskip $

{\bf Acknowledgments:} I am thankful to N. Shnerb, J.Bekenstein, M.Chacham,
S. Oliveira and J. Portnoy for the enlightening conversations.

\end{document}